\documentclass[12pt,letterpaper]{article}
\usepackage{graphicx}
\usepackage{amsmath}

\input{tcilatex}

\begin{document}

\author{V. V. Prosentsov\thanks{%
e-mail: prosentsov@yahoo.com} \\
Stationsstraat 86, 5751 HH, Deurne, The Netherlands}
\title{Super reflector and invisible object: analytical investigation of photonic
cluster reflectivity}
\maketitle

\begin{abstract}
The reflectivity of photonic cluster is important property for nanophotonic
applications especially when one needs to hide the cluster or to make it
extremely visible. Currently, where are no methods clearly describing how to
create the photonic cluster with the predefined reflectivity. In this paper
several analytical  methods are proposed to create the photonic cluster with
minimal or maximal reflectivity. The proposed methods are applicable for the
clusters made of small particles.
\end{abstract}

\section{Introduction}

Invisible men, time traveling, and teleportation are the very known and
fascinating subjects from ancient tales and modern fiction books. For
centuries these subjects were something easy imaginable but ''not of this
world''. While time traveling and teleportation are still far from a
realization (for living organisms at least) the idea of invisibility slowly
began to translate into reality \cite{Hecht}-\cite{Alitalo}. The super
reflection can be considered as antipode to the invisibility and it has many
practical applications already (in Bragg gratings and road signs, for
example). In some sense the invisibility and the super reflection are
closely related but inverted phenomena and that is why they can be studied
in parallel.

The discussion about the invisible objects and nonradiating sources was
started in the scientific literature many years ago (see for example the
works \cite{Dollin}-\cite{Kerker} and the informative review \cite{Gbur}).
Recently it was suggested that metamaterials can be used as building blocks
for invisible objects and the key component of these materials is actually
nanostructured photonic crystal \cite{Gabrielli}-\cite{Alitalo}, \cite
{Metamaterials1}-\cite{Metamater2}. While it was demonstrated that some
devices show relatively low reflectivity in some directions and at some
wavelengths \cite{Alu}, \cite{Alitalo}, the transparent algorithm for design
of invisible photonic clusters is not proposed yet. It is worth to mention
several methods using supernatural conditions. One of them is the
transformation optics method where the coordinates are transformed in such a
way that the object becomes invisible \cite{TransfOpt}-\cite{TransfOpt2}.
The catch of the method is that transformation of the coordinates is
equivalent to the redistribution of the permittivity inside the object and
as the result of such transformation the object becomes invisible. It is not
clear however, how such method can be applied to the photonic crystals
formed by many discrete particles. Another method based on so-called
negative refractive index metamaterials requires supernatural conditions
(see, for example, the review \cite{Negative refract} and the refreshing
work \cite{Negative refract2}) and it is hardly feasible \cite{Litchister}.
Inverse scattering methods \cite{Inverse scatt}-\cite{Inverse scatt2} can be
used, in principle, for construction of invisible objects, however a robust
algorithm is not developed yet.

In this paper I will study visibility of the cluster made of small
dielectric particles by using the local perturbation method (LPM) approach 
\cite{Draine}-\cite{BassVpFr}. I will present and discuss several methods to
construct invisible and extremely visible photonic cluster made of
independent and interacting scatterers.

\section{The field scattered by the photonic cluster in the LPM approximation%
}

The theoretical framework I use is presented in many works (see for example 
\cite{Draine}-\cite{BassVpFr} and references wherein) and it will be only
briefly presented here for convenience and consistency. Consider the
photonic cluster made of particles which characteristic sizes are small
compared to the incident wavelength $\lambda $. For definiteness it is
assumed that the cluster is positioned at the origin of the coordinates. The
electric field $\mathbf{E}$\ propagating in the host medium filled with $N$
small particles is described by the following equation \cite{BassVpFr} 
\begin{equation}
\left( \bigtriangleup -\mathbf{\nabla }\otimes \mathbf{\nabla }+k^{2}\right) 
\mathbf{E}(\mathbf{r})+\frac{k^{2}}{\varepsilon _{0}}\sum_{n=0}^{N-1}\mathbf{%
E}(\mathbf{r}_{n})(\varepsilon _{sc,n}-\varepsilon _{0})f_{n}(\mathbf{r}-%
\mathbf{r}_{n})=\mathbf{S}(\mathbf{r}),  \label{inv11}
\end{equation}
where 
\begin{equation}
k=\frac{2\pi }{\lambda }=\frac{\omega }{c}\sqrt{\varepsilon _{0}},\;f_{n}(%
\mathbf{r}-\mathbf{r}_{n})=\left\{ 
\begin{array}{cc}
1\text{,} & \text{inside particle } \\ 
0\text{,} & \text{outside particle}
\end{array}
\right. .  \label{inv11a}
\end{equation}
Here $\bigtriangleup $ and $\mathbf{\nabla }$\ are the Laplacian and nabla
operators respectively, $\otimes $ defines tensor product, $k\equiv \left| 
\mathbf{k}\right| $ is a wave number in the host medium (the $\left| \mathbf{%
...}\right| $ brackets denote an absolute value), $\omega $ is the angular
frequency, $c$ is the light velocity in vacuum, $\varepsilon _{sc,n}$ and $%
\varepsilon _{0}$ are the permittivity of the $n$-th particle and the medium
respectively, $f_{n}$ is the function describing the shape of the $n$-th
scatterer positioned at $\mathbf{r}_{n}$, and $\mathbf{S}$ is the field
source. The characteristic size of $n$-th scatterer is denoted as $L_{n}$.

It should be noted that the equation (\ref{inv11}) is an approximate one and
it is valid when the small scatterers ($kL_{n}\ll 1$) are considered. The
solution of the equation (\ref{inv11}) is presented in the Appendix A and I
will use in the following discussion the final result presented by the Eq. (%
\ref{a6})

\begin{equation}
\mathbf{E}_{sc}(\mathbf{r})=\frac{k^{2}}{4\pi \varepsilon _{0}}\left( 
\widehat{I}+\frac{\mathbf{\nabla }\otimes \mathbf{\nabla }}{k^{2}}\right)
\sum_{n=0}^{N-1}\mathbf{E}(\mathbf{r}_{n})V_{n}(\varepsilon
_{sc,n}-\varepsilon _{0})\frac{e^{ikR_{n}}}{R_{n}},  \label{inv12}
\end{equation}
where 
\begin{equation}
R_{n}\equiv \left| \mathbf{r}-\mathbf{r}_{n}\right| \gg L_{n}.
\label{inv12a}
\end{equation}
Here $R_{n}$ is the distance between the observer positioned at $\mathbf{r}$
and the $n$-th scatterer placed at $\mathbf{r}_{n}$, $V_{n}$ is the volume
of the $n$-th scatterer.

In many practical cases, the distance between the cluster and the observer
is much more larger than the size of the cluster and the inequality $\left| 
\mathbf{r}\right| \gg \max (\left| \mathbf{r}_{n}\right| )$ is fulfilled. In
this case the scattered field (\ref{inv12}) can be simplified and it can be
rewritten in the far zone ($k\left| \mathbf{r}\right| \gg 1$) in the
following form 
\begin{equation}
\mathbf{E}_{sc}(\mathbf{r})=\frac{k^{2}e^{ikr}}{4\pi r\varepsilon _{0}}%
\left( \widehat{I}-\mathbf{l}\otimes \mathbf{l}\right) \sum_{n=0}^{N-1}%
\mathbf{E}(\mathbf{r}_{n})V_{n}(\varepsilon _{sc,n}-\varepsilon _{0})e^{-ik%
\mathbf{l\cdot r}_{n}},  \label{inv16c}
\end{equation}
where 
\begin{equation*}
\mathbf{l}\equiv \mathbf{r}/r,\;r\equiv \left| \mathbf{r}\right| \gg \max
(\left| \mathbf{r}_{n}\right| )\text{ and }kr\gg 1.
\end{equation*}
The formula (\ref{inv16c}) is the main result of this section and it will be
used extensively in the following discussion. The formula shows that the
field scattered by the cluster is mainly defined by the weighted fields $%
\mathbf{E}(\mathbf{r}_{n})V_{n}(\varepsilon _{sc,n}-\varepsilon _{0})$
inside the small scatterers and by the phase $k\mathbf{l\cdot r}$ in the
scattering direction $\mathbf{l}$. Note, that the interactions between the
particles in the cluster are present in the fields $\mathbf{E}(\mathbf{r}%
_{n})$.

\section{The visibility of the photonic cluster}

\subsection{Criteria of the invisibility and the super reflectivity}

The visibility of any photonic cluster depends on the intensity $I_{sc}(%
\mathbf{r})$ of the field scattered by the cluster (and incident on the
detector positioned at $\mathbf{r}$)\ and on the sensitivity level $\pounds $
of the detector. The sensitivity level of the detector is defined here as
minimal intensity at which a signal is detected. A photonic cluster will be
invisible (in the frequency span $\Delta \omega $ and at the cone of the
directions $\Delta \mathbf{l}$) when $I_{sc}(\mathbf{r})$ is smaller or
equal to $\pounds $. It is convenient to write this statement for the
invisible cluster in the following form 
\begin{equation}
I_{sc}(\mathbf{r})\equiv \left| \mathbf{E}_{sc}(\mathbf{r})\right| ^{2}\leq
\pounds ,\;(\omega \in \Delta \omega \text{ and }\mathbf{l}\in \Delta 
\mathbf{l}),  \label{inv19}
\end{equation}
where $\pounds $ is the sensitivity level of the receiver. The sign $\leq $
in the expression (\ref{inv19}) can be used due to the presence of the noise
masking the detecting signal when $I_{sc}(\mathbf{r})=\pounds $.

To define the super reflectivity one should use the criterium based on the
cluster's reflectivity rather than on the detector sensitivity. The cluster
will be qualified as super reflective when the following relation is true 
\begin{equation}
I_{sc}(\mathbf{r})\sim \Im ,\text{\ }(\omega \in \Delta \omega \text{ and }%
\mathbf{l}\in \Delta \mathbf{l}),  \label{inv19a}
\end{equation}
where $\Im $ is the maximal possible intensity of the field scattered by the
cluster. For the cluster made of small particles the \ maximal scattered
intensity can be estimated with the help of the Eq. (\ref{inv16c}) in the
approximation of the independent particles (when $\mathbf{E}(\mathbf{r}_{n})=%
\mathbf{E}_{in}(\mathbf{r}_{n})/(1+(\varepsilon _{sc,n}-\varepsilon
_{0})\gamma _{n}/3\varepsilon _{0})$) and without depolarization (when$%
\left( \widehat{I}-\mathbf{l}\otimes \mathbf{l}\right) \mathbf{E=}\widehat{I}%
\mathbf{E}$) and it has the form 
\begin{equation}
\Im =\frac{\left| \mathbf{E}_{in}(\mathbf{r})\right| ^{2}k^{4}}{16\pi
^{2}\varepsilon _{0}^{2}r^{2}}\left| \sum_{n=0}^{N-1}V_{n}\epsilon
_{n}\right| ^{2}.  \label{inv19b}
\end{equation}
Here

\begin{equation}
\epsilon _{n}=\frac{\varepsilon _{sc,n}-\varepsilon _{0}}{1+(\varepsilon
_{sc,n}-\varepsilon _{0})\gamma _{n}/3\varepsilon _{0}}  \label{inv19b1}
\end{equation}
where $\gamma _{n}$ is the parameter defined by the shape of the particles
and for spheres, for example, it is 
\begin{equation}
\gamma _{n}=1-k^{2}L_{n}^{2}\left( 1+i\frac{2}{3}kL_{n}\right) ,\;(kL_{n}\ll
1).  \label{inv19ba}
\end{equation}
In the most ideal case one can modify the number of the particles $N$, their
volumes $V_{n}$, positions $\mathbf{r}_{n}$, and the permittivities $%
\varepsilon _{n}$ to make the cluster invisible or extremely visible in
accordance with the conditions (\ref{inv19}) or (\ref{inv19a}) respectively.

\subsection{The ways to achieve the invisibility or super reflectivity}

The scattering process involves the inhomogeneous medium (photonic cluster
in our case) and the incident light. That is why the visibility of the
cluster can be maximized or minimized by the changing the cluster or the
incident light. The second possibility is discussed in the section $7$ where
it is shown how the given cluster can be hidden or made super reflective by
using artificial illumination. In the next section I will discuss approaches
to achieve the invisibility or the super reflectivity of the cluster by
modifying its structure.

The photonic clusters can be subdivided in two main categories: composed of
independent particles and made of interacting particles. The clusters made
of the independent particles will be discussed in the section $5$ while the
clusters composed of the interacting particles will be discussed in the
section $6$.

There are several possibilities to satisfy the conditions (\ref{inv19}) or (%
\ref{inv19a}) and to achieve the invisibility or the super reflectance of
the photonic cluster. One of the ways is to solve the inequations (\ref
{inv19}) and (\ref{inv19a})\ with respect to the positions of the particles $%
\mathbf{r}_{n}$ when the fields $\mathbf{E}(\mathbf{r}_{n})$, the number of
the particles $N$, and the properties of the particles ($\varepsilon _{sc,n}$
and $L_{n}$) are known. (Note that the fields $\mathbf{E}(\mathbf{r}_{n})$
can be found analytically by solving the system of $3N$ linear equations of
type (\ref{a5}).) This way, though clearly formulated, will involve solution
of the system of nonlinear equations that is not feasible. Alternatively,
the positions of the particles in the cluster can be modified till $I_{sc}$
will be smaller than $\pounds $ or of the order of $\Im $. This way was
investigated in the work \cite{Vp} where encouraging results were obtained.
The method closely related to the linear sampling method can be also used.
It will be described in the section $6$ in more detail.

Consider the cluster positioned far from the observer. In this case the
formula (\ref{inv16c}) is applicable and the invisibility condition (\ref
{inv19}) can be presented in the following form 
\begin{equation}
\left| \left( \widehat{I}-\mathbf{l}\otimes \mathbf{l}\right) \mathbf{K}%
\right| \leq \sqrt{\pounds }\frac{4\pi \varepsilon _{0}r}{k^{2}}
\label{inv20}
\end{equation}
and the condition (\ref{inv19a}) for the super reflectivity is 
\begin{equation}
\left| \left( \widehat{I}-\mathbf{l}\otimes \mathbf{l}\right) \mathbf{K}%
\right| \sim \left| \mathbf{E}_{in}(0)\right| \left|
\sum_{n=0}^{N-1}V_{n}\epsilon _{n}\right| .  \label{inv20a}
\end{equation}
Here the vector $\mathbf{K}$ is defined as 
\begin{equation}
\mathbf{K}\equiv \sum_{n=0}^{N-1}\mathbf{E}(\mathbf{r}_{n})V_{n}(\varepsilon
_{sc,n}-\varepsilon _{0})e^{-ik\mathbf{l\cdot r}_{n}}.  \label{inv21}
\end{equation}
The expressions (\ref{inv20}) and (\ref{inv20a}) can be considered as the
equations with respect to the positions $\mathbf{r}_{n}$, $V_{n}$, and $%
\varepsilon _{sc,n}$. The solution of these equations should give the
properties of the particles ($V_{n}$ and $\varepsilon _{sc,n}$) and their
coordinates $\mathbf{r}_{n}$ in the invisible cluster\ or the super
reflector when the observer is positioned at the distance $r$\ from the
cluster in the direction $\mathbf{l}$. The equations (\ref{inv20}) and (\ref
{inv20a}) should be complemented by the additional equations for the fields $%
\mathbf{E}(\mathbf{r}_{n})$ and they should contain at least $N$ equations
formulated for $N$ observation directions $\mathbf{l}$. It should be
emphasized that the equations are nonlinear with respect to the unknowns and
they are extremely complicated. While the equations (\ref{inv20}) and (\ref
{inv20a}) can be solved numerically in principle, a number of serious
drawbacks and limitations exist reducing the value of the numerical
solution. The most serious limitations are the complex values of the
solutions and the excessive number of the solutions.

Below, the conditions of invisibility and super reflectance will be
discussed for the photonic cluster made of small particles.

\section{The zero and the maximal scattering by the cluster of independent
particles}

Consider now the cluster made of $N$ scatterers placed in such a way that
interactions between the particles are negligible. The interaction between a
central particle and $M$ particles surrounding it is negligible when the
cube of the distance between the particles is much larger than the total
volume of the particles, i.e. when the following condition holds

\begin{equation}
R_{mn}^{3}\gg MV_{n}/4.  \label{inv22}
\end{equation}
For more information one can see the formula (\ref{b4}) in the Appendix B
and the required conditions. In this case the\ fields $\mathbf{E}(\mathbf{r}%
_{n})$\ inside the scatterers can be calculated explicitly by using the Eq. (%
\ref{a5}) 
\begin{equation}
\mathbf{E}(\mathbf{r}_{n})=\frac{\mathbf{E}_{in}(\mathbf{r}_{n})}{%
1+(\varepsilon _{sc,n}-\varepsilon _{0})\gamma _{n}/3\varepsilon _{0}},
\label{inv23}
\end{equation}
where $\gamma _{n}$ are the parameters taking into account the
characteristic size $L_{n}$ and the shape of the particles. For example, for
spheres and for cubes $\gamma _{n}$ respectively are 
\begin{eqnarray}
\gamma _{n} &=&1-k^{2}L_{n}^{2}\left( 1+i\frac{2}{3}kL_{n}\right)   \notag \\
&&  \label{inv23a} \\
\gamma _{n} &=&1-k^{2}L_{n}^{2}\left( 1.52+i\frac{4}{\pi }kL_{n}\right) , 
\notag
\end{eqnarray}
where $L_{n}$ is characteristic size of the particle (radius of the sphere
or half of the cube size). Substituting the fields (\ref{inv23}) into the
equations (\ref{inv20}) and (\ref{inv20a})\ we have 
\begin{equation}
\left| \left( \widehat{I}-\mathbf{l}\otimes \mathbf{l}\right)
\sum_{n=0}^{N-1}\mathbf{E}_{in}(\mathbf{r}_{n})V_{n}\epsilon _{n}e^{-ik_{0}%
\mathbf{l\cdot r}_{n}}\right| \leq \sqrt{\pounds }\frac{4\pi \varepsilon
_{0}r}{k^{2}}.  \label{inv23aa}
\end{equation}
\begin{equation}
\left| \left( \widehat{I}-\mathbf{l}\otimes \mathbf{l}\right)
\sum_{n=0}^{N-1}\mathbf{E}_{in}(\mathbf{r}_{n})V_{n}\epsilon _{n}e^{-ik_{0}%
\mathbf{l\cdot r}_{n}}\right| \sim \left| \mathbf{E}_{in}(0)\right| \left|
\sum_{n=0}^{N-1}V_{n}\epsilon _{n}\right| .  \label{inv23b}
\end{equation}
The equations (\ref{inv23aa}) and (\ref{inv23b}) show that even for the
independent particles the cluster's visibility is extremely complex
phenomenon depending on all particles (via the fields $\mathbf{E}(\mathbf{r}%
_{n})$) and it is extremely sensitive to the parameters of each particle ($%
V_{n}$, $\varepsilon _{sc,n}$, $\gamma _{n}$, and $L_{n}$). The equations (%
\ref{inv23aa}) and (\ref{inv23b}) are difficult to solve and that is why
some simplifications are required. Below I will use two important
simplifications: the scatterers with identical permittivity ($\varepsilon
_{sc,n}=\varepsilon _{sc}$) and the long wavelength approximation.

\subsection{The cluster made of independent particles with identical
permittivity}

Consider the case when the particles in the cluster have the same
permittivity $\varepsilon _{sc}$ and the incident field is the plane wave,
i. e. $\mathbf{E}_{in}(\mathbf{r})\equiv \mathbf{A}e^{i\mathbf{k}\cdot 
\mathbf{r}}$. In this case the expressions (\ref{inv23aa}) and (\ref{inv23b}%
) can be essentially simplified and they can be rewritten in the following
ultimate form 
\begin{equation}
\left| \sum_{n=0}^{N-1}V_{n}e^{i\mathbf{k}_{sc}\mathbf{\cdot r}_{n}}\right|
=\left\{ 
\begin{array}{c}
0 \\ 
\sum_{n=0}^{N-1}V_{n}
\end{array}
\right. ,  \label{inv23c}
\end{equation}
where $\mathbf{k}_{sc}=\mathbf{k}-k\mathbf{l}$ is the scattering vector. The
formula (\ref{inv23c}) was obtained in the assumption that the sensitivity
level is $\pounds $ and that depolarization is not essential in this case.
The Eq. (\ref{inv23c}) shows that the visibility of the cluster made of the
independent scatterers is solely defined by the phases $\mathbf{k}_{sc}%
\mathbf{\cdot r}_{n}$ and by the volumes $V_{n}$ playing the role of weight
factors.

The expression (\ref{inv23c}) is actually system of nonlinear equations for
the positions $\mathbf{r}_{n}$. When the positions $\mathbf{r}_{n}$ are
known one can produce invisible or super reflective cluster. The downside is
that even this simplified system (\ref{inv23c}) is difficult to solve
analytically.

However, for some specific systems the solutions are clearly visible.
Consider, for example, the cluster with central symmetry. In this case the
expression (\ref{inv23c}) transforms into the following one

\begin{equation}
\left| \sum_{n=0}^{(N-1)/2}V_{n}\cos (\mathbf{k}_{sc}\mathbf{\cdot r}%
_{n})\right| =\left\{ 
\begin{array}{c}
0 \\ 
\sum_{n=0}^{(N-1)/2}V_{n}
\end{array}
\right. .  \label{inv23c1}
\end{equation}
One set of solutions of the Eq. (\ref{inv23c1})\ is clearly visible and for
the invisibility it is 
\begin{equation}
\mathbf{k}_{sc}\mathbf{\cdot r}_{n}=\pi (1/2+n)  \label{inv23c2}
\end{equation}
and for the maximal reflectivity it is

\begin{equation}
\mathbf{k}_{sc}\mathbf{\cdot r}_{n}=\left\{ 
\begin{array}{c}
2m\pi \\ 
(2m+1)\pi
\end{array}
\right. ,  \label{inv23c3}
\end{equation}
where $m$ is an integer.

It is interesting to note that the maximal scattering happens only when the
conditions (\ref{inv23c3}) are fulfilled and the scattering is not affected
by the ``weight factors'' $V_{n}$. On the other hand, for the minimal
scattering, the weight factors $V_{n}$ are extremely important such that
invisibility can happen in multiple ways (i. e. other solutions are possible
where the weight factors $V_{n}$ play active role).

\subsection{The cluster of independent particles in the long wavelength
approximation}

It is interesting to note that despite the independency (defined by the
condition (\ref{inv22}) the particles in the cluster can be placed such that
the distance between the adjacent scatterers will be much smaller than the
incident wavelength such that the following condition will be satisfied

\begin{equation}
\left| \mathbf{k}_{sc}\mathbf{\cdot }\left( \mathbf{r}_{n}-\mathbf{r}%
_{m}\right) \right| \ll 1,  \label{inv23c6}
\end{equation}
where the particles positioned at $\mathbf{r}_{n}$ and $\mathbf{r}_{m}$ are
the adjacent ones.\ By using the condition (\ref{inv23c6}) the sum (\ref
{inv23aa}) and (\ref{inv23b}) can be replaced by the following integral 
\begin{equation}
\Gamma \equiv \frac{1}{d^{3}}\left| \int_{V_{cl}}\epsilon (\mathbf{r})e^{i%
\mathbf{k}_{sc}\mathbf{\cdot r}}d\mathbf{r}\right| ,  \label{inv23d}
\end{equation}
where 
\begin{equation}
d\equiv \left\langle \left| \mathbf{r}_{n}-\mathbf{r}_{m}\right|
\right\rangle ,\;\epsilon (\mathbf{r})=\frac{(\varepsilon _{sc}(\mathbf{r}%
)-\varepsilon _{0})}{1+(\varepsilon _{sc}(\mathbf{r})-\varepsilon
_{0})/3\varepsilon _{0}}.  \label{inv23d1}
\end{equation}
Here $d$ is the average period of the cluster. Note that the integral $%
\Gamma $\ depends on the shape of the cluster which is not known beforehand
and the shape should be defined from other considerations (practical or
guess, for example).

Since $\epsilon (\mathbf{r})$ vanishes outside of the cluster, the
integration in (\ref{inv23d}) can be extended till infinity and $\Gamma $ is
actually the Fourier transform of the $\epsilon (\mathbf{r})$. In this case
we can present $\Gamma $ as

\begin{equation}
\Gamma =\frac{8\pi ^{3}}{d^{3}}\left| \widetilde{\epsilon }(-\mathbf{k}%
_{sc})\right| ,  \label{inv24}
\end{equation}
where

\begin{equation}
\widetilde{\epsilon }(\mathbf{k}_{sc})\equiv \frac{1}{8\pi ^{3}}%
\int_{-\infty }^{\infty }\epsilon (\mathbf{r})e^{-i\mathbf{k}_{sc}\mathbf{%
\cdot r}}d\mathbf{r}  \label{inv24aa}
\end{equation}
is the Fourier transform of the contrast function $\epsilon (\mathbf{r})$.

The formula (\ref{inv24}) shows that when the long wavelength approximation
is valid, the visibility of the cluster made of the independent scatterers
is defined by the Fourier transform of the contrast function $\epsilon (%
\mathbf{r})$. This property allows to construct the permittivity $%
\varepsilon _{sc}(\mathbf{r})$ of the cluster by using the following formula

\begin{equation}
\varepsilon _{sc}(\mathbf{r})=\varepsilon _{0}\left( 1+\frac{\epsilon (%
\mathbf{r})}{\varepsilon _{0}-\epsilon (\mathbf{r})/3}\right) .
\label{inv24b}
\end{equation}

It is worth to note that this result resembles the one presented in \cite
{Gbur} for the invisible scatterer in the first Born approximation.

\section{The visibility of the cluster of interacting particles: the linear
sampling method approach}

Consider the cluster of arbitrary form made of interacting particles.
Surely, the cluster can be surrounded by the sphere of the radius $\rho $
and we can subdivide the sphere into small cells positioned at the points $%
\mathbf{r}_{n}$ ($\left| \mathbf{r}_{n}\right| \leq \rho $). Suppose that we
know the field scattered by the cluster at the observation points $\mathbf{r}%
_{m}$ such that $\left| \mathbf{r}_{m}\right| =r$. In this case one can
write the following system of equations in respect to the fields $\mathbf{E}(%
\mathbf{r}_{n})$ and the weighted contrasts $\mu _{n}$%
\begin{equation}
\mathbf{E}_{sc}(\mathbf{r}_{m})=\frac{k^{2}e^{ik_{0}r}}{4\pi \varepsilon
_{0}r}\left( \widehat{I}-\mathbf{l}_{m}\otimes \mathbf{l}_{m}\right)
\sum_{n=0}^{N-1}\mathbf{E}(\mathbf{r}_{n})\mu _{n}e^{-ik\mathbf{l}_{m}%
\mathbf{\cdot r}_{n}},  \label{inv25}
\end{equation}
where $m=1...M$ ($M\geq N$) and 
\begin{equation}
\mu _{n}\equiv V_{n}(\varepsilon _{sc,n}-\varepsilon _{0}),\;\mathbf{l}%
_{m}\equiv \mathbf{r}_{m}/r,\;\left| \mathbf{r}_{m}\right| =r.
\label{inv25a}
\end{equation}
The equations (\ref{inv25}) can be resolved in respect to the multiplication
product $\mathbf{E}(\mathbf{r}_{n})\mu _{n}$. The next step is to find the
fields $\mathbf{E}(\mathbf{r}_{j})$ by using the following equations 
\begin{eqnarray}
\mathbf{E}(\mathbf{r}_{j}) &=&\mathbf{E}_{in}(\mathbf{r}_{j})+\mathbf{E}(%
\mathbf{r}_{j})\mu _{j}\gamma _{j}+  \label{inv25b} \\
&&\frac{k^{2}}{4\pi \varepsilon _{0}}\left( a\widehat{I}-b\mathbf{l}%
_{nj}\otimes \mathbf{l}_{nj}\right) \sum_{n\neq j}\mathbf{E}(\mathbf{r}%
_{n})\mu _{n}\frac{e^{ik\left| \mathbf{r}_{n}-\mathbf{r}_{j}\right| }}{%
\left| \mathbf{r}_{n}-\mathbf{r}_{j}\right| },  \notag \\
a &=&1+i/kR-1/k^{2}R^{2}, \\
b &=&-1-3i/kR+3/k^{2}R^{2},
\end{eqnarray}
where $\gamma _{j}$ takes into account the shape of the $j$-th particle and
it is described by the formulae above.

When the field $\mathbf{E}(\mathbf{r}_{j})$ is known, the weighed contrast $%
\mu _{j}$ is fount from the multiplication product $\mathbf{E}(\mathbf{r}%
_{n})\mu _{n}$. The number of the cells ($N$) will be larger than the number
of the particles in the cluster, however the contrast of some cells should
be close to zero such that shape of the cluster (defined by the non zero
contrast) will be not spherical but similar to the real one.

The only open question is the values of the scattered fields $\mathbf{E}%
_{sc}(\mathbf{r}_{m})$ since we need to have definitive values of them to
construct the invisible or the super reflective cluster. We note that the
amplitudes of the fields $\mathbf{E}_{sc}(\mathbf{r}_{m})$ are defined in
some sense by the Eqs. (\ref{inv20}) and (\ref{inv20a}) while the phases are
not clearly defined. This can lead to ambiguity which can be reduced by
imposing additional restrictions (from design or engineering point of view,
for example).

\section{The visibility of the cluster in artificially created incident field%
}

Some applications require to hide a given cluster or make it clearly visible
(super reflective, for example). One of the ways to do it is to synthesize
the incident field $\mathbf{E}_{in}$ in such a way that the cluster will
change its reflectance (at the wavelength $\lambda $ in the direction $%
\mathbf{l}$). To do this, the fields $\mathbf{E}(\mathbf{r}_{n})$ inside the
particles should be found from the following system of equations 
\begin{equation}
\frac{k^{2}e^{ikr}}{4\pi \varepsilon _{0}r}\left( \widehat{I}-\mathbf{l}%
_{m}\otimes \mathbf{l}_{m}\right) \sum_{n=0}^{N-1}\mathbf{E}(\mathbf{r}%
_{n})V_{n}(\varepsilon _{sc,n}-\varepsilon _{0})e^{-ik\mathbf{l}_{m}\mathbf{%
\cdot r}_{n}}=\mathbf{E}_{sc}(\mathbf{r}_{m}),  \label{inv31}
\end{equation}
where $\mathbf{l}_{m}$ and $\mathbf{r}_{m}$\ are the directions and the
points respectively in which the cluster should be invisible. The system (%
\ref{inv31})\ consists of linear equations with respect to the unknown
fields $\mathbf{E}(\mathbf{r}_{n})$ and it can be easily resolved. When the
fields $\mathbf{E}(\mathbf{r}_{n})$ are known, the scattered field $\mathbf{E%
}_{sc}(\mathbf{r}_{n})$ can be found by using the formula (\ref{inv16c}).
The incident fields $\mathbf{E}_{in}(\mathbf{r}_{n})$ are found from the
following formula (see Eq. (\ref{a2}) for the reference) 
\begin{equation}
\mathbf{E}_{in}(\mathbf{r}_{n})=\mathbf{E}(\mathbf{r}_{n})-\mathbf{E}_{sc}(%
\mathbf{r}_{n}).  \label{inv32}
\end{equation}
When the cluster is illuminated by the incident field $\mathbf{E}_{in}$
synthesized in accordance with the expression (\ref{inv32}), the cluster
will be invisible or super reflective for the observers positioned in the
directions $\mathbf{l}_{k}$ at the wavelength $\lambda $.

\section{Conclusions}

The visibility criteria have been discussed for the photonic cluster made of
small particles. The conditions of the zero and of the maximal scattering
have been studied for the clusters made of independent and interacting
particles. It has been shown that in the long wavelength approximation the
visibility of the cluster made of the independent particles is governed by
the Fourier transform of the optical contrast.

The clear algorithm to construct the invisible or the super reflective
photonic cluster has been proposed for the clusters made of interacting
particles.

The new method to hide a given photonic cluster or to make it extremely
visible by creating artificial incident field has been presented.

\begin{equation*}
\end{equation*}

\textbf{Acknowledgments}

I would like to thank Prof. V. Freilikher for important suggestions and
critical comments. Many thanks to my mother Lyudmila for moral support.

\section{Appendix A: The LPM formalism}

The equation (\ref{inv11})

\begin{equation}
\left( \bigtriangleup -\mathbf{\nabla }\otimes \mathbf{\nabla }+k^{2}\right) 
\mathbf{E}(\mathbf{r})+\frac{k^{2}}{\varepsilon _{0}}\sum_{n=0}^{N-1}\mathbf{%
E}(\mathbf{r}_{n})(\varepsilon _{sc,n}-\varepsilon _{0})f_{n}(\mathbf{r}-%
\mathbf{r}_{n})=\mathbf{S}(\mathbf{r}),  \tag{A1}  \label{a1}
\end{equation}
is easily solvable with respect to the fields $\mathbf{E}(\mathbf{r}_{n})$
when the positions $\mathbf{r}_{n}$ of the particles are known (it will
invoke solution of $3N$ linear equations for each frequency $\omega $). The
solution of the equation (\ref{a1}) can be presented in the following form 
\begin{equation}
\mathbf{E}(\mathbf{r})\equiv \mathbf{E}_{in}(\mathbf{r})+\mathbf{E}_{sc}(%
\mathbf{r}),  \tag{A2}  \label{a2}
\end{equation}
where the scattered field $\mathbf{E}_{sc}$ is 
\begin{equation}
\mathbf{E}_{sc}(\mathbf{r})=\frac{k^{2}}{\varepsilon _{0}}\left( \widehat{I}+%
\frac{\mathbf{\nabla }\otimes \mathbf{\nabla }}{k^{2}}\right)
\sum_{n=0}^{N-1}\mathbf{E}(\mathbf{r}_{n})(\varepsilon _{sc,n}-\varepsilon
_{0})\Phi _{n}(\mathbf{r})  \tag{A3}  \label{a3}
\end{equation}
and 
\begin{equation}
\Phi _{n}(\mathbf{r})\equiv \int_{-\infty }^{\infty }\frac{\widetilde{f_{n}}(%
\mathbf{q})e^{i\mathbf{q\cdot (r-r}_{n})}}{(q^{2}-k^{2})}d\mathbf{q,\;}\;%
\widetilde{f_{n}}(\mathbf{q})\equiv \frac{1}{8\pi ^{3}}\int_{-\infty
}^{\infty }f_{n}(\mathbf{u})e^{-i\mathbf{q\cdot u}}d\mathbf{u.}  \tag{A4}
\label{a4}
\end{equation}
Here the incident field $\mathbf{E}_{in}$ is created by the source $\mathbf{S%
}$ in the host medium and it is not important for our consideration (see for
example \cite{Jackson} for more details). The tensor $\widehat{I}$ is the $%
3\times 3$ unitary tensor in polarization space and $\mathbf{r}_{n}$ is the
radius vector of the n-th particle. The field $\mathbf{E}(\mathbf{r}_{n})$
is the field inside the $n$-th particle, $\widetilde{f_{n}}$ is the Fourier
transform of the function $f_{n}$, and $\mathbf{\cdot }$ defines scalar
product. Note that the integration in Eq. (\ref{a4}) is over infinite three
dimensional spaces.

The equations for the fields $\mathbf{E}(\mathbf{r}_{n})$ are found by
substituting $\mathbf{r}=\mathbf{r}_{n}$ into Eq. (\ref{a2}) 
\begin{equation}
\mathbf{E}(\mathbf{r}_{n})=\mathbf{E}_{in}(\mathbf{r}_{n})+\frac{k^{2}}{%
\varepsilon _{0}}\sum_{n=0}^{N-1}(\varepsilon _{sc,n}-\varepsilon
_{0})\int_{-\infty }^{\infty }\frac{\left( \widehat{I}-\frac{\mathbf{q}%
\otimes \mathbf{q}}{k^{2}}\right) \widetilde{f_{n}}(\mathbf{q})}{%
(q^{2}-k^{2})}d\mathbf{q\,\mathbf{E}(\mathbf{r}}_{n}\mathbf{).}  \tag{A5}
\label{a5}
\end{equation}

It should be emphasized that the formula (\ref{a2}) is rather general one
and it describes the field in the medium with photonic cluster of arbitrary
form made of small particles of arbitrary form.

The formula for the scattered field (\ref{a3}) can be simplified when the
distance between the observer and the $n$-th scatterer ($R_{n}$) is large,
i.e. when $R_{n}\gg L_{n}$. In this case the integral $\Phi _{n}$ can be
calculated approximately. Note also that the integral $\Phi _{n}$ can be
calculated exactly at least for the spherical particles. When $R_{n}\gg
L_{n} $ the integration in (\ref{a3}) gives 
\begin{equation}
\mathbf{E}_{sc}(\mathbf{r})=\frac{k^{2}}{4\pi \varepsilon _{0}}\left( 
\widehat{I}+\frac{\mathbf{\nabla }\otimes \mathbf{\nabla }}{k^{2}}\right)
\sum_{n=0}^{N-1}\mathbf{E}(\mathbf{r}_{n})V_{n}(\varepsilon
_{sc,n}-\varepsilon _{0})\frac{e^{ikR_{n}}}{R_{n}},  \tag{A6}
\end{equation}
where 
\begin{equation}
R_{n}\equiv \left| \mathbf{r}-\mathbf{r}_{n}\right| \gg L_{n}.  \tag{A7}
\end{equation}
Here $R_{n}$ is the distance between the observer positioned at $\mathbf{r}$
and the $n$-th scatterer placed at $\mathbf{r}_{n}$, $V_{n}$ is the volume
of the $n$-th scatterer. 
\begin{equation}
\mathbf{E}_{sc}(\mathbf{r})=\frac{k^{2}}{4\pi \varepsilon _{0}}\left( 
\widehat{I}+\frac{\mathbf{\nabla }\otimes \mathbf{\nabla }}{k^{2}}\right)
\sum_{n=0}^{N-1}\mathbf{E}(\mathbf{r}_{n})V_{n}(\varepsilon
_{sc,n}-\varepsilon _{0})\frac{e^{ikR_{n}}}{R_{n}},  \tag{A6}  \label{a6}
\end{equation}
where 
\begin{equation}
R_{n}\equiv \left| \mathbf{r}-\mathbf{r}_{n}\right| \gg L_{n}.  \tag{A7}
\label{a7}
\end{equation}
Here $R_{n}$ is the distance between the observer positioned at $\mathbf{r}$
and the $n$-th scatterer placed at $\mathbf{r}_{n}$, $V_{n}$ is the volume
of the $n$-th scatterer.

\section{Appendix B: The condition of the independence of the particles in
the cluster}

Consider the cluster made of identical small scatterers positioned at $%
\mathbf{r}_{n}$. The cluster consists of the central particle (positioned at 
$\mathbf{r}_{0}$) and the particles surrounding it ($M$ particles are placed
at the distance $R$ from the central particle). The field inside the central
particle positioned at $\mathbf{r}_{0}$ has the following form (see Eq. (\ref
{a5})) 
\begin{equation}
\mathbf{E}(\mathbf{r}_{0})=\left[ \frac{k^{2}e^{ikR}}{4\pi \varepsilon _{0}R}%
V(\varepsilon _{sc}-\varepsilon _{0})\sum_{n=1}^{M}\left( a\widehat{I}+b%
\mathbf{l}_{0n}\otimes \mathbf{l}_{0n}\right) \mathbf{E}(\mathbf{r}%
_{n})+\right.   \tag{B1}  \label{b1}
\end{equation}
\begin{equation*}
\left. \mathbf{E}_{in}(\mathbf{r}_{0})\right] /(1+(\varepsilon
_{sc}-\varepsilon _{0})/3\varepsilon _{0}),
\end{equation*}
where 
\begin{equation}
a=1+i/kR-1/k^{2}R^{2},  \tag{B2}
\end{equation}
\begin{equation*}
b=-1-3i/kR+3/k^{2}R^{2},
\end{equation*}
and 
\begin{equation}
\mathbf{l}_{0n}=\left( \mathbf{r}_{0}-\mathbf{r}_{n}\right) /R,\;R=\left| 
\mathbf{r}_{0}-\mathbf{r}_{n}\right| .  \tag{B3}
\end{equation}
Here $V$ and $\varepsilon _{sc}$ are the volume and the permittivity of the
scatterers respectively.

The importance of the formula (\ref{b1}) is that it allows us to estimate
the distance at which the particles are independent, i.e. when the field
inside the particle located at $\mathbf{r}_{0}$ will be independent from the
particles located at $\mathbf{r}_{n}$. This will happen when the term
containing $\mathbf{E}(\mathbf{r}_{n})$ in Eq. (\ref{b1}) will be much
smaller than the incident field $\mathbf{E}_{in}(\mathbf{r}_{0})$. The
analysis of the Eq. (\ref{b1}) shows that for long wavelengths (when $kR\ll
1 $) the particles are independent when the following condition is satisfied 
\begin{equation}
R^{3}\gg MV/4,  \tag{B4}  \label{b4}
\end{equation}
where $M$ is the number of the particles surrounding the central one. The
condition (\ref{b4}) is used in the paper to discriminate the clusters with
independent and interacting particles. We note that this is over estimated
condition since it was supposed that all the particles interfere in a
constructive way. The condition (\ref{b4}) shows, for example, that for the
cluster with $M=6$ surrounding spheres (case of simple cubic lattice) the
particles will be independent when $R\geq 3.3L$.

\end{document}